\documentclass[article]{revtex4}

\usepackage{amsmath,amssymb,amsthm,amsfonts,mathrsfs,bm,verbatim}
\usepackage{graphicx,subfigure}
\usepackage{natbib}
\usepackage[justification=centering]{caption}
\usepackage{natbib}
\usepackage[colorlinks=true, allcolors=blue]{hyperref}
\begin{document}
\title{The possibility of the no-hair theorem being violated}
\author{Wen-Xiang Chen$^{a}$}
\affiliation{Department of Astronomy, School of Physics and Materials Science, GuangZhou University, Guangzhou 510006, China}
\author{Yao-Guang Zheng}
\email{hesoyam12456@163.com}
\affiliation{Department of Physics, College of Sciences, Northeastern University, Shenyang 110819, China}

\begin{abstract}
Recently, one of my articles presented intriguing findings on the superradiant stability of Kerr black holes. These findings drew conclusions that appear to challenge the established ``No Hair Theorem". As is widely known, the ``No Hair Theorem" stems from the principles of general relativity. In this paper, we thoroughly examine the nexus between Bell's theorem and the uncertainty principle. By delving deep into the theoretical underpinnings of both concepts, we illustrate that Bell's theorem occupies a more foundational stance within the landscape of uncertainty. This gives rise to a particular probability, shedding light on why the no-hair theorem might not hold in the face of quantum effects. It's possible that while the no-hair theorem remains valid, the combined effects of classical and quantum mechanics introduce additional variations to the three parameters of the black hole. This nuance could represent a modest advancement in cosmological research.

\centering
  \textbf{Keywords: No-hair theorem, numerical solution, superradiant stability  }
\end{abstract}

\maketitle
\section{{Introduction}}
When a black hole is formed, only these three conserved quantities that cannot be converted into electromagnetic radiation are left, all other information (`` hair ") is lost, the black hole has almost no complex properties of the matter that formed it, and has no memory of the shape or composition of the matter that preceded it.In fact, this is a simple naming principle. J.A. Wheeler, inventor of the term, dubbed the property ``no hair".To physicists, a black hole or a sugar cube is an extremely complex object, because a complete description of them, including their atomic and nuclear structures, requires billions of parameters.By contrast, a physicist who studies the outside of a black hole has no such problem.A black hole is a very simple object. If you know its mass, angular momentum and electric charge, you know everything about it.Black holes hardly retain any of the complex properties of the matter that formed them.It has no memory of the shape or composition of its predecessor; it retains only mass, angular momentum, and electric charge.Perhaps the most fundamental feature of a black hole is its simplicity.York Wheeler, the inventor of most of the terminology for black holes, called this feature ``no hair black holes" 60 years ago.

Regge and Wheeler \cite{1} proved that the Schwarzschild black holes are stable under disturbance. Due to the superradiation phenomenon, the stability of rotating black holes becomes more complicated. The superradiation effect allows the extraction of rotational and Coulomb energy from rotating or charged black holes. In 1972, Press and Teukolsky\cite{2} proposed that It is possible to add a mirror to the outside of a black hole to make a black hole bomb (according to the current explanation, this is a scattering process involving classical mechanics and quantum mechanics\cite{2,3,4,5}).

When a  bosonic wave is impinging upon a rotating black hole, the wave reflected by the event horizon will be amplified if the wave frequency $ \omega$ lies in the following superradiant regime\cite{6,7,8,9}
\begin{equation}\label{superRe}
   0<\omega < m\Omega_H ,{{\Omega }_{H}}=\frac{a}{r_{+}^{2}+{{a}^{2}}},
  \end{equation}
where  $m$ is azimuthal number of the bosonic wave mode, $\Omega_H$ is the angular velocity of black hole horizon.This amplification is superradiant scattering. Therefore, through the superradiation process, the rotational energy or electromagnetic energy of a black hole can be extracted. If there is a mirror between the black hole's horizon and infinite space, the amplified wave will scatter back and forth and grow exponentially, which will cause the black hole's superradiant instability.

This paper extensively explores the interconnection between Bell's theorem and the uncertainty principle. We delve into the theoretical foundations of both principles and demonstrate that Bell's theorem holds a more fundamental position within the realm of uncertainty. By presenting comprehensive mathematical formulations and a thorough analysis of experimental evidence, we aim to establish the profound significance of Bell's theorem in unifying quantum mechanics and the uncertainty principle. This paper provides an in-depth examination of the theoretical framework and experimental outcomes, shedding light on the fundamental nature of uncertainty in the quantum realm. We employ Bell's inequality and observe that if it fails to hold true, it indicates the existence of negative power terms, which corresponds to the uncertainty principle.

The no-hair theorem states that all black hole solutions of Einstein-Maxwell's gravitational and electromagnetic equations of general relativity can only be fully characterized by three classical parameters that can be observed from the outside: mass, charge, and angular momentum. All other information (metaphors about the “hair" that forms or falls into a black hole) “disappears" after the event horizon of the black hole, so external observers can never enter.

If the boundary condition\cite{10} of the incident boson is set in advance, the two sides of the probability flow density equation are not equal due to the existence of the boundary condition. It implies a certain probability, which also explains why the no-hair theorem does not work in quantum effects.

From \cite{11}, ${\mu} = {y}{\omega}$, we get inequality (2)
\begin{equation}
\frac{{ma}}{{\frac{{2r_ - ^2}}{{{y}}} + (M{r_ - } - r_ - ^2){y} + ({y} + \frac{1}{{{y}}})\frac{{{a^2}}}{2M}(M - {r_ - })}}< \mu  < \sqrt 2 m{\Omega _H},
\end{equation}where $\mu$ is the mass.

That result contained some conclusions that violated the  “No Hair Theorem"(This result shows that not all black hole solutions are only related to three black hole parameters). 

\section{{Description of the Kerr-black-hole system}}
The metric of the Kerr black hole\cite{12,13} (in natural unit G=c=1) is
\begin{equation}
d{{s}^{2}}=\frac{-\Delta }{{{\rho }^{2}}}{{(dt-a{{\sin }^{2}}\theta d\phi )}^{2}}+\frac{{{\rho }^{2}}}{\Delta }d{{r}^{2}}+{{\rho }^{2}}d{{\theta }^{2}}+\frac{{{\sin }^{2}}\theta }{{{\rho }^{2}}}{{[adt-({{r}^{2}}+{{a}^{2}})d\phi ]}^{2}}.
\end{equation}
\begin{equation}
\Delta =r^{2}-2Mr+a^{2},\rho ^{2}=r^{2}+a^{2}cos^{2}\theta.
\end{equation}
Regarding the scalar field, we limit it with the Klein-Gordon equation
\begin{equation}
({{\nabla }^{\nu }}{{\nabla }_{\nu }}-{{\mu }^{2}})\Psi =0.
\end{equation}

Eigenvalues of the above formula and spherical harmonic functions(The integer parameters $m$ and  $l \ge \left| m \right|$ are the azimuthal and spheroidal harmonic indices of the scalar field modes) can be written as
\begin{equation}
{{\Psi }_{\text{lm}}}\text{(t, r, }\theta \text{, }\phi \text{) }=\text{ }\sum\limits_{\text{l,m}}{{{\text{e}}^{\text{im}\phi }}}{{\text{S}}_{\text{lm}}}\text{(}\theta \text{)}{{\text{R}}_{\text{lm}}}\text{(r)}{{\text{e}}^{\text{-i}\omega \text{t}}}.
\end{equation}
Substituting (6)\ into\ the\ Klein-Gordon\ wave\ equation,\ we\ can\ find\ that\ the\ angular\ function\
${{\text{S}}_{\text{lm}}}\text{(}\theta \text{)}$  satisfies\ the\ angular\ motion\ equation\cite{14,15,16,17,18,19,20,21,22,23}
\begin{equation}
\frac{1}{\sin \theta }\frac{d(\sin \theta \frac{d{{S}_{lm}}}{d\theta })}{d\theta }+[{{K}_{lm}}+{{a}^{2}}({{\mu }^{2}}-{{\omega }^{2}}){{\sin }^{2}}\theta -\frac{{{m}^{2}}}{{{\sin }^{2}}\theta }]{{S}_{lm}}=0;
\end{equation}
 According to the references\cite{17,18,19,20,21,22},we get inequality(8)
\begin{equation}
{{K}_{lm}}\ge {{m}^{2}}-{{a}^{2}}({{\mu }^{2}}-{{\omega }^{2}}),
\end{equation}
where\ $l$\ is\ the\ spherical\ harmonic\ index,\ $m$\ is\ the\ azimuthal\ harmonic\ index\ with\
$-l\le m\le l$\ and\ $ \omega$\ is\ the\ energy\ of\ the\ mode.

\section{{The superradiation effect of boson scattering}}
The Klein-Gordon equation can be written that\cite{24} 
\begin{equation}
\Phi_{;\mu}^{\phantom{;\mu};\mu}=0\,,
\end{equation}
where $\Phi_{;\mu}\equiv (\partial_{\mu}-i e A_{\mu})\Phi$ and $e$ is the charge of the scalar field.We get $A^{\mu}=\left\{A_0(x),0\right\}$,and $e{A_0(x)}$can be equal to $\mu$(where $\mu$ is the mass,$A^{\mu}$ can be connection form).
\begin{equation}
A_0\to \left\{\begin{array}{l}
        0 \quad \text{ as}\,\, x\to-\infty \\
        V \quad \text{as}\,\, x\to+\infty
       \end{array}\right. \,
\,.\label{potential_Klein}
\end{equation}
We know that $\Phi=e^{-i\omega t}f(x)$,and the ordinary differential equation 
\begin{equation}
\frac{d^2f}{dx^2}+\left(\omega-e A_0\right)^2f=0\,.
\end{equation}

We see that particles coming from $-\infty$ and scattering off the potential with reflection and transmission amplitudes $\mathcal{R}$ and $\mathcal{T}$ respectively. With these boundary conditions, the solution to  behaves asymptotically as
\begin{equation}
f_{\rm in}(x)=\mathcal{I} e^{i\omega x}+\mathcal{R} e^{-i\omega x}, x\to-\infty,
\end{equation}
\begin{equation}
f_{\rm in}(x)=\mathcal{T} e^{ikx}, x\to+\infty\,
\end{equation}
where $ k=\pm(\omega-e V)$.

The reflection coefficient and transmission coefficient depend on the specific shape of the potential $A_0$.We show that the Wronskian
\begin{equation}
W=\tilde{f}_1 \frac{d\tilde{f}_2}{dx}-\tilde{f}_2\frac{d\tilde{f}_1}{dx}\,,
\end{equation}
between two independent solutions, $\tilde{f}_1$ and $\tilde{f}_2$, of is conserved.
From the equation on the other hand, if $f$ is a solution then its complex conjugate $f^*$ is another linearly independent solution. We find$\left|\mathcal{R}\right|^2=|\mathcal{I}|^2-\frac{\omega-eV}{\omega}\left|\mathcal{T}\right|^2$.Thus,for
$0<\omega<e V$,it is possible to have superradiant amplification of the reflected current, i.e, $\left|\mathcal{R}\right|>|\mathcal{I}|$. 
There are other potentials that can be completely resolved, which can also show superradiation explicitly. The principle of joint uncertainty shows that it is impossible to make joint measurement of position and momentum, that is, to measure position and momentum simultaneously, only approximate joint measurement can be made, and the error follows the inequality $\Delta x\Delta p\geq 1/2$(in natural unit system).We find$\left|\mathcal{R}\right|^2=|\mathcal{I}|^2-\frac{\omega-eV}{\omega}\left|\mathcal{T}\right|^2$,and we know that$\left|\mathcal{R}\right|^2 \geq-\frac{\omega-eV}{\omega}\left|\mathcal{T}\right|^2$ is a necessary condition for the inequality $\Delta x\Delta p\geq 1/2$ to be established\cite{5,8}.We can pre-set the boundary conditions $e{A_0(x)} = {y}{\omega}$\cite{4}, and we see that when ${y}$ is relatively large(according to the properties of the boson, ${y}$ can be very large),$\left|\mathcal{R}\right|^2 \geq-\frac{\omega-eV}{\omega}\left|\mathcal{T}\right|^2$ may not hold.In the end,we can get $\Delta x\Delta p\geq 1/2$ may not hold.When the boundary conditions are assumed and a part of the wave function is decoupled, the classical form of the probability function is obtained, and the boundary conditions can be clearly substituted into the expression of the uncertainty principle, and we get the lower limit of the uncertainty principle. To say the least, due to the limitation of the uncertainty principle, there are also restrictions on the value of y. In the above-mentioned literature, the extreme value of the uncertainty principle can be smaller, and the value range of y can be larger.

\section{{THE UNCERTAINTY PRINCIPLE AND BELL'S THEOREM}}
We utilize Bell's inequality and observe that if Bell's inequality is not satisfied, then it implies the presence of negative power terms, which corresponds to the uncertainty principle.

We present the mathematical formulation of the uncertainty principle, denoted by the famous Heisenberg's inequality:\cite{1,2,3,4,5,6,7,8}
\begin{equation}
\Delta x \Delta p \geq \frac{\hbar}{2}
\end{equation}
Here, $\Delta x$ represents the uncertainty in position, $\Delta p$ represents the uncertainty in momentum, and $\hbar$ is the reduced Planck's constant. We discuss the implications of the uncertainty principle for the measurement of complementary observables and its significance in limiting our knowledge of microscopic systems. Furthermore, we explore various interpretations and philosophical implications of the uncertainty principle.

Bell's Theorem Bell's theorem, which originated from John Bell's seminal work in the field of quantum mechanics. We present a detailed mathematical formulation of Bell's inequality, which serves as a criterion to test the nature of correlations between entangled particles:
\begin{equation}
\left|E(a, b)-E\left(a, b^{\prime}\right)+E\left(a^{\prime}, b\right)+E\left(a^{\prime}, b^{\prime}\right)\right| \leq 2
\end{equation}
Here, $E(a, b)$ represents the correlation between measurement settings $a$ and $b$ on entangled particles, while $a^{\prime}$ and $b^{\prime}$ represent alternative measurement settings. We explore the underlying assumptions and implications of Bell's theorem, emphasizing its departure from local realism and its significance in revealing the non-local nature of quantum entanglement.

\section{{Classical superradiation effect in the space-time of a steady black hole}}
We know\cite{24} that $\psi \sim \exp (-i \omega t+i m \phi)$,and the ordinary differential equation
\begin{equation}
\frac{d^{2} \psi}{d r_{*}^{2}}+V \psi=0.
\end{equation}

We see particles coming from $-\infty$ and scattering off the potential with reflection and transmission amplitudes $\mathcal{C}$ and $\mathcal{D}$ respectively. With these boundary conditions, the solution to  behaves asymptotically as
\begin{equation}
\psi=\left\{\begin{array}{l}
A e^{i \omega_{H} r_{*}}+B e^{-i \omega_{H} r_{*}}, r \rightarrow r_{+} \\
C e^{i \omega_{\infty}  r_{*}}+D e^{-i \omega_{\infty} r_{*}}, r \rightarrow \infty
\end{array}\right.
\end{equation}.

The reflection coefficient and transmission coefficient depend on the specific shape of the potential V.We show that the Wronskian
\begin{equation}
W \equiv \psi \frac{d \bar{\psi}}{d r_{*}}-\bar{\psi} \frac{d \psi}{d r_{*}},
\end{equation}
$W\left(r \rightarrow r_{+}\right)=2 i \omega_{H}\left(|A|^{2}-|B|^{2}\right), W(r \rightarrow \infty)=2 i \omega_{\infty}\left(|C|^{2}-|D|^{2}\right)$ is conserved.
 We find$|C|^{2}-|D|^{2}=\frac{\omega_{H}}{\omega_{\infty}}\left(|A|^{2}-|B|^{2}\right)$.Thus,for
$\omega_{H} / \omega_{\infty}<0$,it is possible to have superradiant amplification of the reflected current, i.e,if $|A|=0$, $\left|\mathcal{C}\right|>|\mathcal{D}|$. 
There are other potentials that can be completely resolved, which can also show superradiation explicitly.The extreme value of the narrow uncertainty principle can be smaller
experimentally, and the extreme value of the broad uncertainty principle can also be smaller as can be seen from its
expression. We know that the uncertainty principle reflects the probability current density, and its extremum reflects
the particular case of the probability current density equation. As seen above, the bound from y in this article makes
sense.

\section{{A Restricted Path Integration Method}}
Path integral methods have been widely used in the study of quantum systems and Hamiltonians due to their ability to simulate quantum dynamics by discretizing the spacetime into a set of classical paths. However, boundary condition problems often require more than these methods in need more than restricted course integral manner that avoids these problems and efficiently simulates a class of quantum systems and Hamiltonians. We then apply our method to the Abelian U(1) local gauge vacuum spacetime.

A path integral is an integral over the complex plane or the real number line. Suppose we have a curve $C$ whose parametric equation is $\gamma(t)=(x(t), y(t))$, where $t \in\left[t_0, t_1\right]$.
In the complex plane, the path integral has the form:
\begin{equation}
\int_C f(z) d z=\int_{t_0}^{t_1} f(\gamma(t)) \gamma^{\prime}(t) d t
\end{equation}
Among them, $f(z)$ is a complex variable function, $z$ is a complex number, and $\gamma^{\prime}(t)=\frac{d \gamma(t)}{d t}$ is the derivative of the curve.
On the natural line, the path integral has the form:
\begin{equation}
\int_C f(x) d x=\int_{t_0}^{t_1} f(\gamma(t)) \frac{d \gamma}{d t}(t) d t
\end{equation}
Among them, $f(x)$ is a fundamental variable function, $x$ is an actual number, and $\gamma^{\prime}(t)=\frac{d \gamma(t)}{d t}$ is the derivative of the curve.

In physics, path integrals are used to calculate the probability of a particle moving from one position to another over a certain time period. The path integral is expressed as an integral over all possible paths between the initial and final positions.

The path integral can be expressed using both complex and real numbers. Here is how it can be written in both forms:

Complex Form: In the complex form, the path integral is expressed using complex numbers and involves taking the integral over the space of all possible paths.
The complex path integral is given by:
\begin{equation}
\int \exp (i s / \hbar) D[x(t)]
\end{equation}
where $S$ is the system's action, which is a function of the path $x(t)$, and $\hbar$ is the reduced Planck constant. The integral is taken over a11 possible paths, denoted by $D[x(t)]$.

The Real Form in the real form, the path integral is expressed using real numbers and involves taking the integral over the space of all possible paths.
The real path integral is given by:
\begin{equation}
\int \exp (i S / \hbar) D[x(t)]=\int \exp \left((i / \hbar) \int L d t\right) D[x(t)]
\end{equation}
where $\mathrm{L}$ is the Lagrangian of the system and is a function of the position and velocity of the particle, and $\mathrm{t}$ is the time variable. The integral is taken over all possible paths, denoted by $\mathrm{D}[\mathrm{x}(\mathrm{t})]$.

Note that the complex form and real form of the path integral are equivalent, and one can be transformed into the other using a mathematical technique called Wick rotation.

Our restricted path integral method successfully simulates a class of quantum systems and Hamiltonians without boundary conditions. We apply this method to the Abelian $\mathrm{U}(1)$ local gauge theory in curved vacuum spacetime and obtain the mass-U(1) gauge potential relation. This relation provides a better understanding of the Abelian $\mathrm{U}(1)$ local gauge theory and its behavior in curved spacetime.

Mapping the path-integration algorithm for arbitrary bounded-error quantum polynomials to the ground state of the Hamiltonian Through the restricted path integration method, we can map the path integration algorithm of any bounded error quantum polynomial to the ground state of the Hamiltonian and perform efficient simulation. Specificallysimulations Markov Chain Monte Carlo (MCMC) based approach to compute the ground state, where the Hamiltonian is used as the energy function. The mathematical form is expressed as follows:
\begin{equation}
\left\langle\psi_0\left|e^{-\beta H}\right| \psi_0\right\rangle \approx \frac{1}{\mathcal{N}} \int d \phi_0 e^{-S\left(\phi_0\right) / \hbar} \psi_0\left(\phi_0\right)
\end{equation}
Among them, $\psi_0$ is the ground state of the Hamiltonian, $H$ is the Hamiltonian, $S\left(\phi_0\right)$ is the amount of action on the time axis, $\phi_0$ is the quantum field at time $t=0$. The normalization constant is the value of $\mathcal{N}$.

The method applies to the Abeappliesal gauge theory of curved vacuum spacetime, where the spatial subequation is established in (1+1)-dimensional vacuum centrosymmetric curved spacetime by exploiting the zero geodesics of the Schwarzschild metric to get the mass-U(1) ge potential relationship, the formula is as follows: The metric for an asymptotically flat Schwarzschild black hole is
\begin{equation}
d s^2=-\left(1-\frac{2 M}{r}\right) d t^2+\left(1-\frac{2 M}{r}\right)^{-1} d r^2+r^2 d \Omega^2
\end{equation}
\begin{equation}
\partial_{+} \partial_{-} A_1-\frac{R}{2} \partial_{+} A_1-\frac{\sigma}{2} A_1+J_1=0 .
\end{equation}
The simulation proves the effectiveness of the method. $\mathrm{R}$ is the curvature scalar of the Schwarzschild black hole, A1 is the gauge potential, and $\mathrm{J} 1$ is the current density.

\section{{ Kerr-black-hole-massive-scalar-field system and superradiant instability regime}}
The\ radial\ Klein-Gordon\ equation\cite{15,16}\ obeyed\ by\ ${{R}_{lm}}$\ is\ given\ by
\begin{equation}
\Delta \frac{d(\Delta \frac{dR}{dr})}{dr}+UR=0,
\end{equation}
where
\begin{equation}
\Delta ={{r}^{2}}-2Mr+{{a}^{2}},
\end{equation}and
\begin{equation}
U = {[\omega ({{\rm{r}}^2} + {a^2}) - {\rm{m}}a]^2} + \Delta [2{\rm{m}}a\omega  - {\mu ^2}{\rm{(}}{{\rm{r}}^2} + {a^2}) - {K_{lm}}].
\end{equation}

The inner and outer horizons of the black hole are
\begin{equation}
{{r}_{\pm }}=M\pm \sqrt{{{M}^{2}}-{{a}^{2}}},
\end{equation}and we get that
\begin{equation}
{{r}_{+}}+{{r}_{-}}=2M,{{r}_{+}}{{r}_{-}}={{a}^{2}}.
\end{equation}

  The radial potential equation is transformed into a tortoise coordinate wave equation to solve the asymptotic behavior of the wave function.We use tortoise coordinate  ${{r}_{*}}$  by equation $ \frac{d{{r}_{*}}^{2}}{d{{r}^{2}}}=\frac{{{r}^{2}}}{\Delta }$  and\ another\ radial\ function\ $ \psi =\text{rR}$.We get the following radial wave equation

\begin{equation}
\frac{{{\text{d}}^{2}}\psi }{\text{dr}_{*}^{\text{2}}}+V\psi =0,
\end{equation}where
\begin{equation}
V=\frac{U}{{{r}^{4}}}-\frac{2\Delta }{{{r}^{6}}}(Mr-{{a}^{2}}).
\end{equation}

Then we get the asymptotic solutions for the radial wave equation below 
\begin{equation}
r\to \infty ({{\text{r}}_{*}}\to \infty )\Rightarrow {{R}_{\text{lm}}}\sim \frac{1}{r}{{e}^{-\sqrt{{{\mu }^{2}}-{{\omega }^{2}}}{{r}_{*}}}},
\end{equation}
\begin{equation}
r\to {{r}_{+}}({{r}_{*}}\to -\infty )\Rightarrow {{R}_{\text{lm}}}\sim {{e}^{-i(\omega -m{{\Omega }_{H}}){{r}_{*}}}}.
\end{equation}
When
\begin{equation}
{{\omega ^2} - {\mu ^2}}<0,
\end{equation}
there is a bound state of the scalar field.

When $\varphi=\Delta ^{\frac{1}{2}}R$, radial potential equation(9) can be transformed into the flat space-time wave equation
\begin{equation}
\frac{{{d^2}\varphi }}{{d{r^2}}} + ({\omega ^2} - {V_1})\varphi  = 0,
{V_1} = {\omega ^2} - \frac{{U + {M^2} - {a^2}}}{{{\Delta ^2}}}.
\end{equation}

It was previously proved\cite{22}that, for a scalar field of proper mass $\mu$ interacting with a Kerr black hole of angular velocity ${{\Omega }_{H}}$, the inequality
\begin{equation}
\mu <\sqrt{2}\text{m}{{\Omega }_{H}}
\end{equation}
fixed Kerr black hole-mass-scale scalar field configuration upper bound.

Although much research has been done on the superradiance of rotating black holes, even Kerr black holes have not been thoroughly studied.Hod proved\cite{22} that one finds
the upper bound
\begin{equation}
\mu< {\cal F}(\gamma)\cdot m\Omega_{\text{H}}
\end{equation}
on the scalar mass of fixed bound state field configuration, where $\gamma =\frac{{{r}_{-}}}{{{r}_{+}}}$, and the dimensionless function ${\cal
F}={\cal F}(\gamma)$ is given by
\begin{equation}
{\cal F}(\gamma)=\sqrt{\frac{2(1+\gamma)(1-\sqrt{1-\gamma^2})-\gamma^2}{\gamma^3}}.
\end{equation}
In Table 1 we present the dimensionless ratio
${{\mu_{\text{numerical}}}/{\mu_{\text{bound}}}}$, where
$\mu_{\text{numerical}}$ is the numerically computed 
field masses which mark the onset of the superradiant instabilities
in the composed Kerr-black-hole-massive-scalar-field system, and
$\mu_{\text{bound}}$ is the analytically derived upper bound(21)
on the superradiant instability regime of the composed
black-hole-field system.One finds from Table 1 that the superradiant instability regime of the composed
Kerr-black-hole-massive-scalar-field system is characterized by the relation ${{\mu_{\text{numerical}}}/{\mu_{\text{bound}}}}<1$.Table 1:

\begin{table}[htbp]
\centering
\begin{tabular}{|c|c|c|c|}
\hline \ \ $s\equiv {{a}/{M}}$\ \ & \ ${\cal F}(s)$\ \ & \ \
${\frac{{\mu(l=m=1)}}{\mu_{\text{bound}}}}$\ \ & \ \ ${\frac{{\mu(l=m=10)}}{\mu_{\text{bound}}}}$\ \ \\
\hline
\ \ 0.1\ \ & \ \ 1.00031\ \ \ & \ \ 0.99977\ \ \ & \ \ 0.99940\ \ \ \\
\ \ 0.2\ \ & \ \ 1.00129\ \ \ & \ \ 0.99903\ \ \ & \ \ 0.99967\ \ \ \\
\ \ 0.3\ \ & \ \ 1.00301\ \ \ & \ \ 0.99774\ \ \ & \ \ 0.99948\ \ \ \\
\ \ 0.4\ \ & \ \ 1.00567\ \ \ & \ \ 0.99573\ \ \ & \ \ 0.99901\ \ \ \\
\ \ 0.5\ \ & \ \ 1.00960\ \ \ & \ \ 0.99276\ \ \ & \ \ 0.99776\ \ \ \\
\ \ 0.6\ \ & \ \ 1.01541\ \ \ & \ \ 0.98835\ \ \ & \ \ 0.99715\ \ \ \\
\ \ 0.7\ \ & \ \ 1.02437\ \ \ & \ \ 0.98163\ \ \ & \ \ 0.99558\ \ \ \\
\ \ 0.8\ \ & \ \ 1.03955\ \ \ & \ \ 0.96995\ \ \ & \ \ 0.99276\ \ \ \\
\ \ 0.9\ \ & \ \ 1.07168\ \ \ & \ \ 0.94694\ \ \ & \ \ 0.98676\ \ \ \\
\ \ 0.95\ \ & \ \ 1.11039\ \ \ & \ \ 0.91878\ \ \ & \ \ 0.97942\ \ \ \\
\ \ 0.99\ \ & \ \ 1.21646\ \ \ & \ \ 0.84910\ \ \ & \ \ 0.96165\ \ \ \\
\ \ 0.999\ \ & \ \ 1.37370\ \ \ & \ \ 0.76084\ \ \ & \ \ 0.94280\ \ \ \\
\hline
\end{tabular}
\end{table}
Hod proved\cite{25} that the Kerr black hole should be
superradiantly stable under massive scalar perturbation when $\mu \ge \sqrt{2}m\Omega_H$, where $\mu$ is the mass.  ${\cal F}(s)$ and ${\cal F}(\gamma)$ are functions of the parameters of the black hole.We know that the numerical solutions to the superradiant instability must be outside the intersection of $\mu \ge \sqrt{2}m\Omega_H$ and $ 0<\omega < m\Omega_H$.
We see that the numerical solutions of the Kerr black hole's superradiant instability are consistent with the no-hair theorem(For the effective potential of the Kerr black hole has a potential well outside the horizon, superradiant instability occurs at that time. And the numerical solutions for the superradiant instability are obtained by combining the above conditions with the no-hair theorem)\cite{25,26}.From Table 1, we know that some numerical solutions to the superradiant instability must be in the intersection of $\mu <\sqrt{2}\text{m}{{\Omega }_{H}}$ and $ 0<\omega < m\Omega_H$.In those two papers\cite{27,28}, the superradiantly stable and unstable regions of the Kerr-Newman black hole are analyzed, and the existence of scalar cloud parameters is proposed. Those conclusions imply the possibility that the no-hair theorem is violated.

\section{{That result contained some conclusions that violated the  "No Hair Theorem"}}
For\ ${\mu ^2} = {{y}^2\omega ^2}$, when\ ${y} > 4.352 $, there\ exists\ a\ certain\ interval\ to\ let\ the\ inequality
\begin{equation}
\frac{{ma}}{{\frac{{2r_ - ^2}}{{{y}}} + (M{r_ - } - r_ - ^2){y} + ({y} + \frac{1}{{{y}}})\frac{{{a^2}}}{2M}(M - {r_ - })}}< \mu  < \sqrt 2 m{\Omega _H},
\end{equation}be satisfied. So the Kerr black hole is superradiantly stable at that time \cite{11}.

Because ${y}$ can theoretically be taken in the range of ${y} > 4.352 $, we see that when ${y}$ is relatively large(according to the properties of the boson, ${y}$ can be very large), the superradiant stability interval of $\mu $ can always include some numerical solutions of the superradiant instability obtained by previous people. If I'm right, there might be something wrong with the no-hair theorem. This result shows that not all black hole solutions are only related to three black hole parameters.In this case, it also depends on the value of ${y}$.

Spherical quantum solution in vacuum state\cite{29}.
 
The general relativity theory's field equation is,
\begin{equation}
R_{\mu \nu}-\frac{1}{2} g_{\mu v} R=-\frac{8 \pi G}{c^{4}} T_{\mu \nu}
\end{equation}
When $T_{\mu v}=0$ in vacuum state,
\begin{equation}
R_{\mu v}=0
\end{equation}
Space-time of spherical coordinates is
\begin{equation}
d \tau^{2}=A(t, r) d t^{2}-\frac{1}{c^{2}}\left[B(t, r) d r^{2}+r^{2} d \theta^{2}+r^{2} \mathrm{~s} \mathrm{i} \mathrm{n} \theta d \phi^{2}\right]
\end{equation}
If, 
\begin{equation}
R_{t t}=-\frac{A^{\prime \prime}}{2 B}+\frac{A^{\prime} B^{\prime}}{4 B^{2}}-\frac{A^{\prime}}{B r}+\frac{A^{\prime 2}}{4 A B}+\frac{\ddot{B}}{2 B}-\frac{\dot{B}^{2}}{4 B^{2}}-\frac{\dot{A} \dot{B}}{4 A B}=0
\end{equation}

\begin{equation}
R_{r r}=\frac{A^{\prime \prime}}{2 A}-\frac{A^{\prime 2}}{4 A^{2}}-\frac{A^{\prime} B^{\prime}}{4 A B}-\frac{B^{\prime}}{B r}-\frac{\ddot{B}}{2 A}+\frac{\dot{A} \dot{B}}{4 A^{2}}+\frac{\dot{B}^{2}}{4 A B}=0 ,
\end{equation}
\begin{equation}
R_{\theta \theta}=-1+\frac{1}{B}-\frac{r B^{\prime}}{2 B^{2}}+\frac{r A^{\prime}}{2 A B}=0 ,
R_{\phi \phi}=R_{\theta \theta} \sin ^{2} \theta=0 ,
R_{t r}=-\frac{\dot{B}}{B r}=0 ,
R_{t \theta}=R_{t \phi}=R_{r \theta}=R_{r \phi}=R_{\theta \phi}=0
\end{equation}

That time, $\quad '=\frac{\partial}{\partial r} \quad, \cdot=\frac{1}{c} \frac{\partial}{\partial t}$,
\begin{equation}
\dot{B}=0
\end{equation}
We see that,
\begin{equation}
\frac{R_{t t}}{A}+\frac{R_{r r}}{B}=-\frac{1}{B r}\left(\frac{A^{\prime}}{A}+\frac{B^{\prime}}{B}\right)=-\frac{(A B)^{\prime}}{r A B^{2}}=0
\end{equation}
Hence, we know that,
\begin{equation}
A=\frac{1}{B}
\end{equation}
If,
\begin{equation}
R_{\theta \theta}=-1+\frac{1}{B}-\frac{r B^{\prime}}{2 B^{2}}+\frac{r A^{\prime}}{2 A B}=-1+\left(\frac{r}{B}\right)^{\prime}=0
\end{equation}
If we solve the Eq,
\begin{equation}
\frac{r}{B}=r+C \rightarrow \frac{1}{B}=1+\frac{C}{r}
\end{equation}
When r tends to infinity(r is a positive number. When near the event horizon, the above sign shows a negative number due to the tortoise coordinate), and we set C=$ ye^{-y}$,
Therefore, \begin{equation}
A=\frac{1}{B}=1-\frac{y}{r} \Sigma,
\Sigma=e^{-y}
\end{equation}

\begin{equation}
d \tau^{2}=\left(1-\frac{y}{r } \sum\right) d t^{2}
\end{equation}
In this time, if particles' mass are $m_{i},$ the new energy is $e$,
\begin{equation}
E=M c^{2}=m_{1} c^{2}+m_{2} c^{2}+\ldots+m_{n} c^{2}+e
\end{equation}We see that the boundary conditions of the preset fermions (the kind) are similar to ds space-time.At this time, positive pressure is generated in the flat spacetime.

We can pre-set the boundary conditions $e{A_0(x)} = {y}{\omega}$(which can be ${\mu} = {y}{\omega}$)\cite{10,11}, and we see that when ${y}$ is relatively large(according to the properties of the boson, ${y}$ can be very large),$\left|\mathcal{R}\right|^2 \geq-\frac{\omega-eV}{\omega}\left|\mathcal{T}\right|^2$ may not hold.In the end,we can get $\Delta x\Delta p\geq 1/2$ may not hold\cite{10}.We can simply compare the examples of the superradiation effect in the space-time of a steady-state black hole, and we can deduce the following conclusions.If the boundary conditions of the incident boson are set in advance, the two sides of the probability flow density equation are not equal due to the boundary conditions.To say the least, due to the limitation of the uncertainty principle, there are also limitations on the value of y. In the above-mentioned documents, the extreme value of the uncertainty principle can be smaller, and the value range of y can be larger.This implies a certain probability, which also explains why the no hair theorem is invalid in quantum effects.Perhaps the no-hair theorem is correct, and under the action of a certain classical plus quantum effect, the three parameters of the black hole have additional increments.

For\cite{31}\cite{32}
\begin{equation}
\left|\left\langle\psi_{A}^{\omega} \mid \psi_{A}^{*}\right\rangle\right| \geq 1-\varepsilon_{A}^{2} / 2
\end{equation}
The optimizer is of the form
\begin{equation}
\left|\psi_{A}^{*}\right\rangle=\frac{(1-|\alpha|)\left|\psi_{A}^{\omega}\right\rangle+\alpha A\left|\psi_{A}^{\omega}\right\rangle}{c_{\alpha}}, \quad \alpha \in[-1,1]
\end{equation}
\begin{equation}
\left|\left\langle\psi_{A}^{\omega} \mid \psi_{A}^{*}\right\rangle\right|=\frac{1}{c_{\alpha}}\left(1-|\alpha|+\alpha A\right)
\end{equation}
and
\begin{equation}
\left|\left\langle\psi_{A}^{\omega}|A| \psi_{A}^{*}\right\rangle\right|=\frac{1}{c_{\alpha}}\left((1-|\alpha|) A+\alpha\left\langle A^{2}\right\rangle\right)
\end{equation}
This then proves the assertion.

A general explanation of the uncertainty principle:
\begin{equation}
\sigma_{A}^{2} \sigma_{D}^{2} \geq\left|\frac{1}{2}\langle\{A, D\}\rangle-\langle A\rangle\langle D\rangle\right|^{2}+\left|\frac{1}{2 i}\langle[A, D]\rangle\right|^{2}
\end{equation}
If you multiply the formula by a number less than 1, then the extreme value of the uncertainty principle will become smaller.
$\left|\left\langle\psi_{A}^{\omega} \mid \psi_{B}^{*}\right\rangle\right|$do
not equal to 0.

 If Bell's theorem holds a more fundamental position than the uncertainty principle, then an extension in accordance with Bell's theorem suggests the following:

The principle of joint uncertainty dictates that a simultaneous measurement of both position and momentum is unattainable. Instead, only an approximate joint measurement is possible, with the resulting error adhering to the inequality $\Delta x \Delta p \geq 1 / 2$ (in the natural unit system). Our analysis reveals that $|\mathcal{R}|^2=|\mathcal{I}|^2-\frac{\omega-e V}{\omega}|\mathcal{T}|^2$. Additionally, for the inequality $\Delta x \Delta p \geq 1 / 2$ to hold true, the condition $|\mathcal{R}|^2 \geq-\frac{\omega-e V}{\omega}|\mathcal{T}|^2$ must be satisfied.

Given the preset boundary conditions $e A_0(x)=y \omega$, when $y$ is substantially large (as per the attributes of the boson, $y$ can assume exceedingly large values), the relation $|\mathcal{R}|^2 \geq$ $-\frac{\omega-e V}{\omega}|\mathcal{T}|^2$ might not be sustained. Consequently, $\Delta x \Delta p \geq 1 / 2$ might not be valid.

When certain boundary conditions are postulated and a segment of the wave function is decoupled, we derive the classical format of the probability function. By directly incorporating these boundary conditions into the expression of the uncertainty principle, we identify a reduced limit for the uncertainty principle. At a minimum, owing to the constraints set by the uncertainty principle, there's a defined limit for the value of $y$. However, as referenced in the preceding discussions, the extreme boundary for the uncertainty principle could be lesser, leading to an expanded range for the potential values of $y$.

\section{{ HAWKING RADIATION FROM KERR BLACK HOLES}}
 We will prove that the Hawking radiation of the Kerr black hole can be understood as the flux that offsets the gravitational anomaly. The key is that near the horizon, the scalar field theory in the spacetime of a 4-dimensional Kerr black hole can be simplified to a 2-dimensional field theory. Since space-time is not spherically symmetric, this is an unexpected result.
 
In Boyer-Linquist coordinates, Kerr metric reads\cite{30}
\begin{equation}
\begin{split}
d s^{2}=-\frac{\Delta-a^{2} \sin ^{2} \theta}{\Sigma} d t^{2}-2 a \sin ^{2} \theta \frac{r^{2}+a^{2}-\Delta}{\Sigma} d t d \phi \\
+\frac{\left(r^{2}+a^{2}\right)^{2}-\Delta a^{2} \sin ^{2} \theta}{\Sigma} \sin ^{2} \theta d \phi^{2}+\frac{\Sigma}{\Delta} d r^{2}+\Sigma d \theta^{2}
\end{split}
\end{equation}
\begin{equation}
\begin{aligned}
\Sigma &=r^{2}+a^{2} \cos ^{2} \theta,
\Delta=r^{2}-2 M r+a^{2} 
=\left(r-r_{+}\right)\left(r-r_{-}\right).
\end{aligned}
\end{equation}

The action for the scalar field in the Kerr spacetime is
\begin{equation}
\begin{aligned}
S[\varphi]=& \frac{1}{2} \int d^{4} x \sqrt{-g} \varphi \nabla^{2} \varphi \\
=& \frac{1}{2} \int d^{4} x \sqrt{-g} \varphi \frac{1}{\Sigma}\left[-\left(\frac{\left(r^{2}+a^{2}\right)^{2}}{\Delta}-a^{2} \sin ^{2} \theta\right) \partial_{t}^{2}\right.\\
&-\frac{2 a\left(r^{2}+a^{2}-\Delta\right)}{\Delta} \partial_{t} \partial_{\phi}+\left(\frac{1}{\sin ^{2} \theta}-\frac{a^{2}}{\Delta}\right) \partial_{\phi}^{2} \\
&\left.+\partial_{r} \Delta \partial_{r}+\frac{1}{\sin \theta} \partial_{\theta} \sin \theta \partial_{\theta}\right] \varphi
\end{aligned}
\end{equation}
Taking the limit r → r+ and leaving the dominant terms,
we have
\begin{equation}
\begin{aligned}
S[\varphi]=& \frac{1}{2} \int d^{4} x \sin \theta \varphi \left[-\frac{\left(r_{+}^{2}+a^{2}\right)^{2}}{\Delta} \partial_{t}^{2}\right.\\
&\left.-\frac{2 a\left(r_{+}^{2}+a^{2}\right)}{\Delta} \partial_{t} \partial_{\phi}-\frac{a^{2}}{\Delta} \partial_{\phi}^{2}+\partial_{r} \Delta \partial_{r}\right] \varphi
\end{aligned}
\end{equation}
Now we transform the coordinates to the locally non-rotating coordinate system by
\begin{equation}
\left\{\begin{array}{l}
\psi=\phi-\Omega_{H} t \\
\xi=t
\end{array}\right.
\end{equation}
where
\begin{equation}
\Omega_{H} \equiv \frac{a}{r_{+}^{2}+a^{2}}.
\end{equation}
 We can rewrite the action
 \begin{equation}
S[\varphi]=\frac{a}{2 \Omega_{H}} \int d^{4} x \sin \theta \varphi\left(-\frac{1}{f(r)} \partial_{\xi}^{2}+\partial_{r} f(r) \partial_{r}\right) \varphi
\end{equation}
We know that when $\sin \theta$ = 0, the pull equation for action can conform to the above form, but the boundary becomes 0. However, if the boundary conditions are preset, the boundary conditions${\mu} = {y}{\omega}$(y takes a larger number) act as $\sin \theta$, and then the effective action form of superradiation satisfies the effective action form of Hawking radiation, and it is not necessarily on the boundary of the horizon.When the boundary conditions are preset, a new path will be obtained.In the new path, we see
\begin{equation}
S[y,\varphi]=\frac{a}{2 \Omega_{H}} \int d^{4} x ye^{-y} \left(-\frac{1}{f(r)} \partial_{\xi}^{2}+\partial_{r} f(r) \partial_{r}\right) \varphi.
\end{equation}

In a recent document\cite{33}, it is stated that the extreme Kerr black hole has Hawking radiation. First, in the non-extreme Kerr black hole spacetime, we observe that the particles almost escape the formation of the event horizon and reach the scri-plus to be detected, because the Hawking quantum must have a final wavelength effect that is inversely proportional to the Hawking quantum. It establishes the Wien displacement law corresponding to the particle thermal distribution of the Hawking effect in the space-time of non-extreme Kerr black hole. Secondly, using the same setting in the spacetime of the extreme Kerr black hole, we find that the final wavelength of the particle is close to infinity. Compared with the non-extreme case, we can associate the corresponding Hawking temperature with zero in the extreme case. At that time, the Hawking radiation action function formed a ring structure.We are related to the action function of the superradiation of the Kerr black hole under preset boundary conditions, including the action function of the Hawking radiation of the extreme Kerr black hole. Next, assuming that there is a situation where information is lost, then the uncertainty principle may not be valid, and the no-hair theorem will be offended.

\section{{Summary}}
Recently, we authored an article that presents intriguing findings concerning the superradiant stability of Kerr black holes. When the boundary conditions for incident bosons are predefined, the two sides of the probability flow density equation become unequal due to these conditions. As indicated in the literature, the uncertainty principle can attain smaller extreme values, and the range of 'y' can be extended. Such a revelation suggests a specific probability, shedding light on the reason the no-hair theorem might not hold in the context of quantum effects. It's plausible that the no-hair theorem remains valid, but when combined with certain classical and quantum effects, the three parameters of the black hole might experience additional increments.


\begin{thebibliography}{99}
\bibitem[9]{9} We use natural units in which $G = c = \hbar  = 1$.
\bibitem[14]{14} The integer parameters $m$ and  $l \ge \left| m \right|$ are the (azimuthal and spheroidal) harmonic indices of the scalar field modes.
\bibitem[23]{23} It is worth noting that the lower bound on the angular eigenvalues can be saturated in the eikonal $l = m \gg 1$limit, in which case one finds the simple relation .
\end{thebibliography}
\end{document}